\documentclass[aip,reprint]{revtex4-1}
\usepackage[dvips]{graphicx}
\usepackage{amsmath}

\begin{document}

\title{Field-assisted Shockley-Read-Hall recombinations in III-Nitride quantum wells}

\author{Aurelien David}
\email{adavid@soraa.com}
\affiliation{Soraa Inc., 6500 Kaiser Dr. Fremont CA 94555 USA}
\author{Christophe A. Hurni}
\affiliation{Soraa Inc., 6500 Kaiser Dr. Fremont CA 94555 USA}
\author{Nathan G. Young}
\affiliation{Soraa Inc., 6500 Kaiser Dr. Fremont CA 94555 USA}
\author{Michael D. Craven}
\affiliation{Soraa Inc., 6500 Kaiser Dr. Fremont CA 94555 USA}

\date{\today}

\begin{abstract}
The physical process driving low-current non-radiative recombinations in high-quality III-Nitride quantum wells is investigated. Lifetime measurements reveal that these recombinations scale with the overlap of the electron and hole wavefunctions and show weak temperature dependence, in contrast with common empirical expectations for Shockley-Read-Hall recombinations. A model of field-assisted multiphonon point defect recombination in quantum wells is introduced, and shown to quantitatively explain the data. This study provides insight on the high efficiency of III-Nitride light emitters.
\end{abstract}


\maketitle

Studies of the efficiency of III-Nitride light emitting diodes (LEDs) often focus on their high-current properties; insight into the physics of low-current and peak efficiency remains scarce. In particular, the demonstration of very efficient LEDs, despite the strong suppression of radiative recombination by polarization fields,\cite{Waltereit00} remains puzzling. Besides, the low-current regime is of crucial importance for basic physical understanding and for future application such as micro-LED displays. 

In this Letter, we study the physics of non-radiative recombination (NRR) in the low-injection regime --before the onset of current droop-- in high-efficiency III-Nitride quantum well (QW) structures. With all-optical differential carrier lifetime measurements, we show that these recombinations display a strong dependence on the carrier wavefunctions in the QW and weak temperature dependence. These effects are quantitatively explained with a theory of field-assisted recombinations at point defects, driven by the polarization field across the QW.

We first briefly review previous studies. In experiments, NRR is interpreted in the framework of Shockley-Read-Hall (SRH) theory where the NRR rate reads $G_{NR}=An$, with $n$ the carrier density and $A$ an SRH coefficient whose value is fitted empirically, with reported values\cite{Schiavon13,Pristovsek17} of $10^5-10^8$~s$^{-1}$ -- a wide range often explained away by invoking varying material quality. The current-dependence of the internal quantum efficiency is commonly described in the $ABC$ model, with $B$ the radiative coefficient and $C$ interpreted as an Auger coefficient.

On the theoretical front, first-principle studies have exposed the potential role of point defects as NRR centers and have computed corresponding SRH rates.\cite{Alkauskas14,Wikramaratne16} Yet, these calculations pertain to bulk materials, whereas the most relevant effect for LEDs is NRR in the QW active region. In device modeling, investigations are usually carried out in the drift-diffusion framework, using a constant and empirical SRH coefficient. Some device modeling studies have focused on the contribution of transport effects to low-current efficiency: they have shown the importance of trap-assisted tunneling processes (including free carriers tunneling across the junction and confined carriers escaping from the QW), which account for the high ideality factor often reported in III-Nitride LEDs at low current.\cite{Aufdermaur14,Mandurrino15,Desanti16} Yet, there as well, NRR inside the QW itself is still treated with empirical SRH coefficients.

We begin this work with a commentary on the efficiency of III-Nitride QW LEDs. It is well-known that the polarization field across the QW causes a separation of the electron and hole wavefunctions ($\psi_e$, $\psi_h$) and therefore a decrease of the overlap $I=\int{\psi_e \psi_h dz}$. This in turn reduces the oscillator strength and the radiative coefficient $B$, which should scale with $I^2$. Therefore, one might expect that efficiency would depend dramatically on $I$, due to the competition with NRR channels. Yet, this stands in contrast to our experimental observations, which call for a NRR theory beyond the customary use of empirical SRH coefficients. 

\begin{figure}[!!!tttth]
\includegraphics[width=8.5cm]{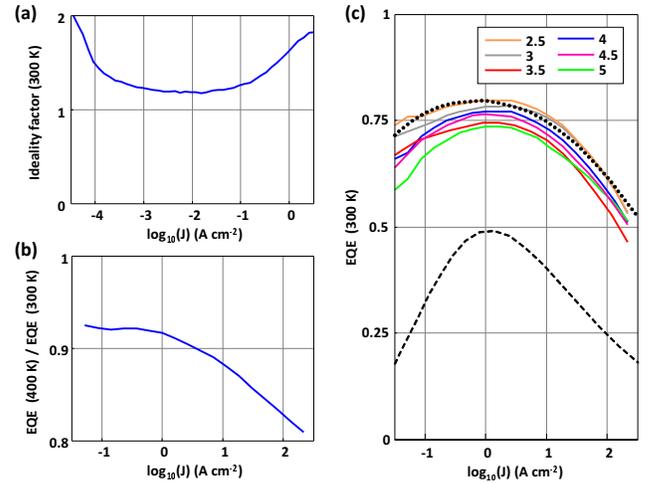}
\caption{(a) Ideality factor of the 4 nm SQW LED. (b) Ratio of $EQE$ at 300~K to 400~K, for the 4 nm SQW LED. (c) Color lines: External quantum efficiency of SQW LEDs of varying thickness as shown in the legend. Black lines: $ABC$ model predictions; dotted line: $ABC$ fit of the $2.5$ nm sample; dashed line: $ABC$ model for a sample of same $A$, while $B$ and $C$ are reduced by 100$\times$. }
\label{Fig_EQE}
\end{figure}

To illustrate this, we grew a series of single-quantum-well (SQW) LEDs on c-plane bulk GaN substrates with a dislocation density $\sim 10^6$~cm$^{-2}$. The SQWs have an indium composition of $10\%$ and their width is varied from 2.5~nm to 5~nm. Importantly, these LEDs display textbook-like ideality factors (Fig.~\ref{Fig_EQE}(a)). This is consistent with Ref.~\onlinecite{David16b}, in which we showed that LEDs on bulk GaN do not display the anomalously-high ideality factors characteristic of heteroepitaxial GaN LEDs.\cite{Aufdermaur14} Likewise, the hot-cold ratio of external quantum efficiency (EQE) remains high ($\sim 0.92$) at low current (Fig.~\ref{Fig_EQE}(b)); again, this differs from defective heteroepitaxial LEDs where a low hot-cold ratio is observed.\cite{Desanti16} Therefore, in our samples, the low-current efficiency is dictated by NRR inside the QW rather than by transport-related effects such as those discussed in Refs.~\onlinecite{Aufdermaur14,Mandurrino15,Desanti16}. This enables the study of NRR physics without extrinsic effects.

Notably, all these LEDs are highly efficient, with a peak EQE of $77\% \pm 3\%$ (Fig.~\ref{Fig_EQE}(c)) and only a minor decrease in efficiency from the thinnest to the thickest SQW, despite the large expected difference in oscillator strength (more than 100$\times$). If we were to derive $ABC$ parameters for the 2.5 nm sample, then predict the EQE of the 5 nm sample by assuming $I^2$ is reduced hundredfold (with a commensurate reduction in $B$, and also $C$ following Refs.~\onlinecite{David10b,Kioupakis12,David17a}) while $A$ remains constant, we would expect a peak EQE of only $50\%$ for that sample, as shown on Fig.~\ref{Fig_EQE}(c). In contrast to such predictions, experimental data indicate that the \textit{relative magnitude} of the radiative and SRH rates must remain approximately constant across these samples. This effect was  tentatively argued in Refs.~\onlinecite{David10b,Kioupakis12}, but has so far lacked solid experimental observation or theoretical proof.

\begin{figure*}[!!!thhhhhhhhhhhhb]
\includegraphics[width=\textwidth]{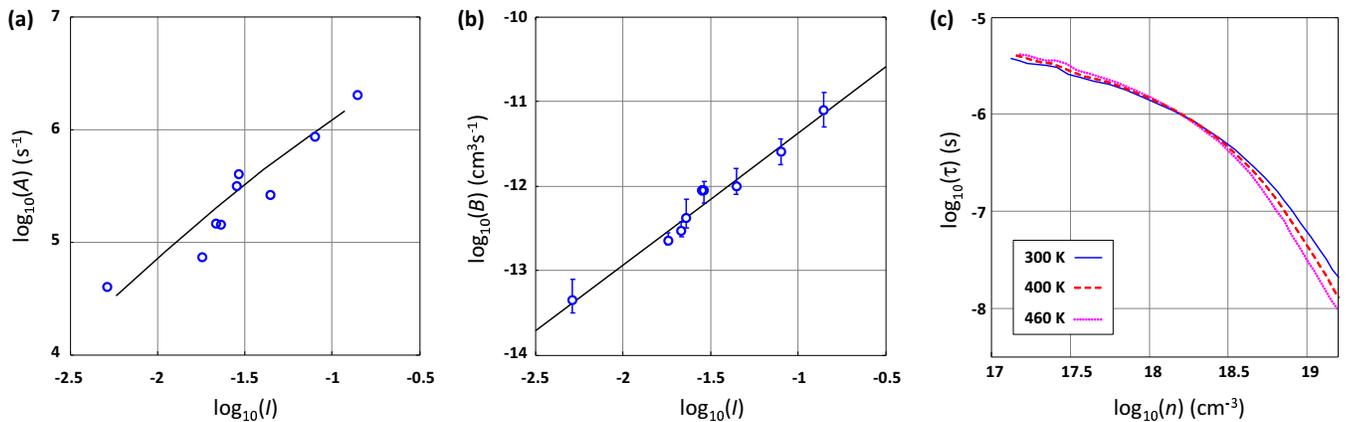}
\caption{Optical differential lifetime results. (a) SRH coefficient $A$ versus wavefunction overlap $I$. Circles: experimental data. Line: model prediction. (b) Radiative coefficient $B$ versus $I$. Circles: experimental data evaluated at $n=10^{18}$cm$^{-3}$. Error bars: span of $B$ for $n$ in the range  $3 \times 10^{17}$-- $3 \times 10^{18}$cm$^{-3}$. Line: best linear fit. (c) Differential lifetime $\tau$ of the 4 nm-thick SQW at various temperatures: the low-current lifetime is nearly constant.}
\label{Fig_data}
\end{figure*}

To clarify the relationship between wavefunctions and recombinations, we turn to an all-optical differential carrier lifetime (ODL) measurement.\cite{David17a} Unlike standard electroluminescent-based measurements, ODL doesn't involve carrier transport/capture effects,\cite{David16a} and therefore provides the recombination lifetime without ambiguity. This is especially important at low current for the study of NRR, since the true SRH lifetime is accessed.

The samples measured in ODL are a second series of SQWs, similar to the LEDs described above but with a thicker intrinsic region optimized for ODL measurements.\cite{David17a} Under optical excitation, the induced photobias of $2$-$3$~V yields a band structure similar to electrical injection,\cite{David10c} while remaining in open-circuit. Table~\ref{table} gives the samples' details: some have a constant In concentration of $\sim 11\%$ (as determined by X-ray diffraction) and varying thickness, while others have a varying In content and a constant thickness of 4 nm. In this moderate range of In $\%$, we expect that point defect incorporation is approximately constant across samples.

\begin{table}%[H] add [H] placement to break table across pages
\caption{\label{table} Details of ODL samples.}
\begin{ruledtabular}
\begin{tabular}{c c c | c c c}
t (nm)     &  In \% & log$_{10}(I)$ & t (nm)& In \% & log$_{10}(I)$ \\
\hline
2.5 & 10.1  &  -0.85 &   4 & 10.7 &  -1.64 \\
3 & 10.9    &  -1.10 &   4 & 11.2 &  -1.67 \\
3.5 & 10.7  &  -1.35 &   4 & 12.7 &  -1.74 \\
4 & 9.0     &  -1.54 &   5 & 10.7 &  -2.29 \\
4 & 9.2     &  -1.55 &     &      & \\
\end{tabular}
\end{ruledtabular}
\end{table}

The ODL measurement gives experimental access to the carrier density $n$ and to the radiative and non-radiative recombination rates ($G_R$, $G_{NR}$). The NRR coefficient $A$ is extracted from $G_{NR}=A n$. We find that, for all samples, $G_{NR}/n$ reaches a well-defined plateau at low current, such that $A$ is unambiguously determined.

As discussed in Ref.~\onlinecite{David17a}, the radiative dynamics are more complex. The radiative rate should read $G_R = B n^2$; however, experimentally $B$ displays an additional weak carrier dependence at low density (before the onset of carrier screening), which we attribute to the impact of carrier localization on radiative recombinations. Nonetheless, a low-current value of $B$ can be derived, within an uncertainty range caused by this $n$-dependence.

Fig.~\ref{Fig_data} shows the experimental values of $A$ and $B$, plotted versus the wavefunction overlap $I$. $A$ and $B$ vary by two orders of magnitudes across the series of samples. The values of $A$ correspond to SRH lifetimes as slow as tens of $\mu$s, testifying to the high quality of the samples. 

Note that $I$ is computed from a one-dimensional Schrodinger solver\cite{Bandeng} which ignores further overlap effects cause by in-plane carrier localization. Indeed we assume that the full wavefunctions $\Psi$ can approximately be factorized (i.e. $\Psi=\psi (z) \times \phi(x,y)$), and that the in-plane term $\phi$ doesn't vary strongly across our samples, so that variations in its overlap contribution can be discarded.

$B$ shows an experimental power-dependence $B\sim I^p$. The value of $p$ slightly depends on the carrier density at which $B$ is evaluated: $p=1.6 \pm 0.3$. This is in reasonable agreement with the theoretical expectation of $p=2$, especially given our simplified calculation of $I$.

Crucially, $A$ also displays a power dependence $I^q$, with $q \sim 1.2$. Thus, the NRR rate effectively scales with the wavefunction overlap, with an exponent somewhat similar to that of the radiative rate. This explains how the LEDs of Fig.~\ref{Fig_EQE} maintain nearly-constant peak efficiency: the significant variations in $B$ are largely compensated by similar variations in $A$. To second order, since $B$ has a slightly larger dependence than $A$ on $I$, the low-current efficiency (equal to $Bn/A$) is moderately reduced in samples with lower $I$ -- as can be seen on Fig.~\ref{Fig_EQE}(c).

Intuitively, this dependence on $I$ is plausible since a point defect must interact with both an electron and a hole for NRR to take place. Interestingly, the dependence of $A$ and $B$ on $I$ holds whether variations in $I$ are caused by a change in QW thickness or composition (at least in the modest composition range spanned here). Further, although the topic of this article is not high-current recombinations, it is worth mentioning that the high-current non-radiative coefficient $C$ (not shown) likewise shows a similar dependence on $I$, already discussed in Refs.~\onlinecite{David10b,Kioupakis12,David17a}.

We note that various previous studies of $A$ have failed to observe the dependence on $I$ displayed in Fig.~\ref{Fig_data}(a), instead reporting large scatter across samples.\cite{Schiavon13,Pristovsek17} Following Ref.~\onlinecite{Reklaitis17}, we hypothesize that the values of $A$ in some of those reports may in fact characterize carrier escape from the QW\cite{Vahala89} and transport-related recombinations in defective heteroepitaxial LEDs,\cite{Aufdermaur14,Mandurrino15,Desanti16} rather than intrinsic NRR in the QW. Besides, a variation in $A$ can in general be caused both by a change in wavefunction overlap and in material quality; in some cases these two effects counteract each other, for instance when the In content is increased significantly (which can deteriorate material quality while increasing the QW field). Thus a low variation in A cannot unambiguously be interpreted as constant material quality.\cite{Nippert16b} Ignoring such considerations can lead to misleading conclusions on the magnitude of NRR.

Fig.~\ref{Fig_data}(c) shows the temperature dependence of the lifetime for the 4~nm sample: from 300~K to 460~K, it is nearly-constant at low current (corresponding to a relative variation in $A$ of less than 10\%). This stands in contrast to other studies where a more significant temperature dependence was observed,\cite{Desanti16, Nippert16b} and to a possible empirical expectation that $A$ may have exponential temperature dependence (based on the exponential temperature-dependence often observed in capture cross-sections\cite{Henry77}).

In summary, our notable experimental findings include the absolute value of $A$, its strong dependence on wavefunction overlap, and its weak temperature dependence.

To explain the behavior of $A$, we now introduce a model of field-assisted point-defect recombinations in the QW (or, equivalently, of defect-assisted tunneling across the QW). This process is a form of the field-assisted multiphonon recombination often observed in semiconductor devices,\cite{Henry77,Hurkx92,Zheng94,Palma97,Mandurrino15} but with QW-quantized states participating in the recombination.

\begin{figure}[!!!tttth]
\includegraphics[width=8.5cm]{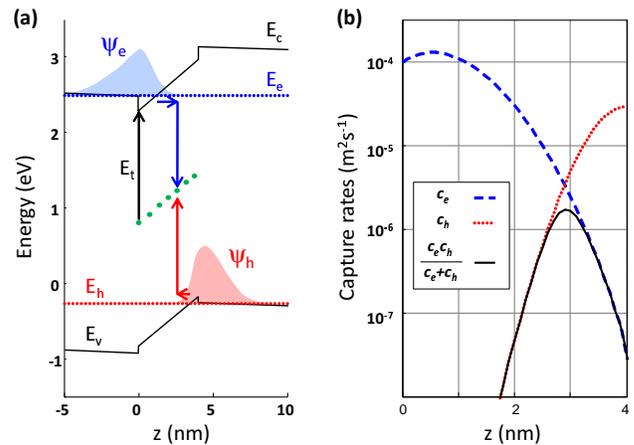}
\caption{(a) Band structure of the 4 nm sample, and illustration of the field-assisted recombination. The wavefunctions penetrate towards the center of the QW, and can recombine at various trap positions. (b) Values of the capture coefficients $c_e, c_h$ and $c_e c_h/(c_e+c_h)$ across the QW. The latter is strongly peaked, at a position where most NRR occur.}
\label{Fig_process}
\end{figure}

A recombination cycle consists of two captures (an electron and a hole) by the same trap (Fig.~\ref{Fig_process}(a)). Due to the polarization field across the QW, the wavefunctions $\psi_e$ and $\psi_h$ are separated, but their decaying tails penetrate towards the center of the QW. Both carriers can thus reach a given point defect, enabling capture.

We introduce standard notations of multiphonon theory, with subscripts $e$ for electrons and $h$ for holes: $\hbar \omega$ is the energy of the phonon mode involved in the capture, $S$ the corresponding Huang-Rhys factor, $\bar{n}$ the phonon occupation number, $\Delta E=E_e-E_t$ (with $E_t$ the trap energy and $E_e$ the quantized energy level), $P=\Delta E /\hbar \omega$ the average number of transition phonons, $a=\hbar/(2 m_e \Delta E)^{1/2}$ the Bohr radius of a trapped electron of effective mass $m_e$, $\xi=2 S [\bar{n}(\bar{n}+1)]^{1/2}$ and $\chi=(P^2+\xi^2)^{1/2}$. The following quantities vary spatially: $\Delta E$, $P$, $a$, and $\chi$.

The interaction potential $V$ coupling a trap to a free state has been derived for various models of the trap wavefunction.\cite{Zheng94,Palma97} We follow Ref.~\onlinecite{Zheng94}, modified to account for the free carrier's presence probability at the trap.\cite{Palma97,Alkauskas14} Since $a \sim 1 \AA$, the quantized electron's wavefunction $\psi_e$ is approximately constant across the extent of the trap and $V$ reads: $\left| V \right| ^2 = 16 \pi a^3 S (\hbar \omega)^2 \left| \psi_e \right|^2 / \Sigma$ (with $\Sigma$ the crystal area). The capture coefficient by a point defect located at position $z$ ensues:

\begin{equation}
\label{eq:c_e}
c_e = \frac{16 \pi^2 a^3}{\hbar}(\Delta E-S \hbar \omega)^2 \left| \psi_e(z) \right|^2 G(\Delta E, T) F(T)
\end{equation}

Here $c_e$ is a surface capture coefficient, expressed in m$^2$~s$^{-1}$. $c_h$ is defined similarly for holes. Eq.~\ref{eq:c_e} involves the auxiliary function:

\begin{equation}
\label{eq:G}
G = \frac{1}{\hbar \omega \sqrt{2 \pi \chi} } \left( \frac{\xi}{P+\chi} \right)^{P} \exp\left( \chi- (2\bar{n}+1)S +\frac{\Delta E}{2 k T} \right)
\end{equation}

Finally, $F(T)$ is a charge-dependent function ($e.g.$ $F>1$ for an attractive center). For simplicity we ignore charge effects such as the change of trap charge when one carrier type is captured (such effects would require additional hypotheses, and would only slightly affect the magnitude of $c_e$, $c_h$) . We therefore assume neutral defects and, following Refs.~\onlinecite{Zheng94,Kang96}, we set $F=1$.

Following a standard derivation,\cite{Abakumov91} the SRH rate for a slice of material of thickness $dz$ is:

\begin{equation}
\label{eq:dG}
dG_{NR}=N_t \frac{c_e n \times c_h p}{c_e n + c_h p} dz ,
\end{equation}

where we have neglected the background-density terms, which are small for deep traps in a wide bandgap. In ODL measurements, $n=p$. The net NRR rate is obtained by integration across $z$, and finally takes the form $G_{NR}=An$, where $A$ is the effective SRH coefficient:

\begin{equation}
\label{Eq:A_SRH}
A = \int{N_t\frac{c_e c_h}{c_e+c_h}dz}
\end{equation}

Eq.~\ref{Eq:A_SRH} resembles the standard bulk SRH expression, but contains wavefunction terms within $c_e$ and $c_h$. We also note that the standard SRH form is recovered because $n=p$; in a general case where $n \neq p$ however, Eq.~\ref{eq:dG} would not integrate to a familiar bulk SRH expression of the form $G_{NR} \sim c_e n \times c_h p/(c_e n + c_h p)$.

We now seek to account for the experimental results quantitatively, proceeding by numerical evaluation of Eq.~\ref{Eq:A_SRH}. We assume a uniform defect density $N_t$.\footnote{This assumption is not crucial: the integrals are heavily peaked, so that $N_t$ needs only be approximately constant near the center of the QW for our results to hold} The model's free parameters are $N_t$ (which merely scales the absolute value of $A$), the trap energy $E_t$, and the electron-phonon parameters $S$ and $\hbar \omega$. 

Evaluating Eq.~\ref{Eq:A_SRH}  for the samples of Table~\ref{table}, we observe that, regardless of the free parameters' values (within plausible ranges), $A$ approximately follows a power law $A\sim I^q$, in general accordance with our experimental findings (the value of $q$ depends on $E_t$, $S$ and $\hbar \omega$). This result is non-trivial given the expressions of $A$ and $I$; we believe it is tied to the rapid decay of the wavefunctions in the QW, which causes integrals such as $I$ and $A$ to be dominated by the value of $\psi_e \times \psi_h$ near the QW center.

Good experimental fits can be obtained with a variety of parameter values -- typically, for $E_t$ in the range $1$-$2$~eV (corresponding to deep traps) and values of $S \hbar \omega$ on the order of 1~eV (a realistic value for a strongly-coupled defect\cite{Alkauskas14}). As an example, we consider a mid-gap state ($E_t=1.6$~eV from the conduction band) with a moderate density $N_t=1.5 \times 10^{14}$ cm$^{-3}$, $\hbar \omega=90$~meV (the energy of the free LO-phonon, which is of the correct order of magnitude) and $S=11$. 

Fig.~\ref{Fig_process}(b) shows the corresponding values of $c_e$, $c_h$ and $c_e c_h/(c_e+c_h)$ across the QW. The latter quantity is strongly peaked at a position near the QW center (with an offset mostly caused by the heavier hole mass): this is where SRH recombination predominantly occurs. This implies that contrary to what is sometimes suggested, NRR is unlikely at the QW/barrier interfaces, at least when the present wavefunction-driven process is relevant.

The resulting fit to experimental data is shown on Fig.~\ref{Fig_data}(a). The absolute magnitude of $A$ and its effective dependence on $I$ are both accurately reproduced. These fitting parameters correspond to a bulk SRH coefficient on the order of $10^7$ s$^{-1}$ (consistent with the extrapolated intersect of the data of Fig.~\ref{Fig_data}(a) with the bulk limit $I=1$).

In addition, for these parameters, $A$ shows a very low temperature sensitivity (not shown): it increases by 10\% from 300~K for 460~K, again in agreement with experimental data. This weak dependence is caused by several factors. First, the high-temperature behavior of $c_e$ and $c_h$ generally scales with $exp[-E_B/kT]$;\cite{Henry77} here the activation energy $E_B=(E_t-S\hbar\omega)^2/4S\hbar\omega$ is low ($\sim$100~meV) due to the large relaxation energy $S\hbar\omega$. Second, in such a low-barrier regime, the non-exponential prefactors of $G$ become relevant and further reduce the thermal sensitivity (in some cases, multiphonon recombination can even lead to a decrease of the capture rate with temperature, as measured in Ref.~\onlinecite{Henry77} and predicted in Ref.~\onlinecite{Zheng94}). Finally, the energy of LO phonons in GaN is high (several $kT$), such that the phonon population $\bar{n}$ is actually not in the high-temperature regime, even at 460K.

We end this Letter with a few implications of our findings. First, we propose that the wavefunction-dependence of the SRH rate plays a fundamental role for the efficiency of III-Nitride LEDs. Crucially, this explains a long-standing puzzle in this field: how c-plane LEDs can maintain high efficiency despite the strong polarization fields which separate wavefunctions. Besides, this clarifies why a variety of LED structures can achieve high efficiency; why experimental values of $A$ vary so widely across studies; and why SRH recombinations are still observed in semi-polar and non-polar devices despite a large increase in oscillator strength. Second, the derived expression for the SRH rate can be easily integrated into device simulators; we hope this will lead to more predictive models and will in turn improve the understanding of III-Nitride device physics. For instance, this may benefit the active field of wavefunction-engineering, which often seeks to increase $B$ without concern for the related effect on $A$.\cite{Zhao11} Finally, this NRR process is expected to be of relevance in any device where electron-hole wavefunctions are significantly separated, including polar materials and type-II heterostructures.

In conclusion, we have identified a process explaining low-current non-radiative recombinations in high-quality III-Nitride QWs. These recombinations show a strong dependence on carrier wavefunctions overlap and a weak temperature dependence. A field-assisted multiphonon recombination process accounts for this data: wavefunctions are separated by polarization fields but penetrate towards the center of the QW where point-defect recombination occurs. This effect can cause non-radiative recombinations to vary by several orders of magnitude independently from material quality, and is of fundamental importance for understanding the efficiency of 	Nitride LEDs: it competes with the well-known wavefunction-dependence of radiative recombinations, explaining why the efficiency of these LEDs is relatively insensitive to wavefunction overlap, and why c-plane LEDs have historically shown high efficiency despite their strong polarization fields.

%\bibliography{Biblio_These}

\end{document}